\documentclass[
 preprint,footinbib,
 amsmath,amssymb,
 aps,prb,showpacs, showkeys
]{revtex4-1}
\usepackage[utf8]{inputenc}
\usepackage{graphicx}
\DeclareGraphicsExtensions{.png}
\usepackage{dcolumn}
\usepackage{bm}
\usepackage{url}
\usepackage{amsmath}
\usepackage{float}
\usepackage{caption}
\usepackage{subcaption}
\let\oldAA\AA
\renewcommand{\AA}{\text{\normalfont\oldAA}}

\begin{document}

\title{The higher superconducting transition temperature T$_c$ and the functional derivative of T$_c$ with $\alpha^2F(\omega)$ for electron-phonon superconductors}

\author{J. A. Camargo-Martínez}
\email{jcamargo@unitropico.edu.co}
\affiliation{Grupo de Investigación en Ciencias Básicas, Aplicación e Innovación - CIBAIN, Unitrópico, Yopal, Casanare, Colombia}

\author{G. I. González-Pedreros}
\affiliation{Grupo de Investigación en Ciencias Básicas, Aplicación e Innovación - CIBAIN, Unitrópico, Yopal, Casanare, Colombia}

\author{F. Mesa}
\affiliation{Grupo NanoTech, Facultad de Ciencias Naturales y Matemáticas, Universidad del Rosario, Cra. 24 No. 63C-69, Bogotá, Colombia}

\date{\today}

\begin{abstract}

This work presents an analysis of the functional derivative of the superconducting transition temperature T$_c$ with respect to the electron–phonon coupling function $\alpha^2F(\omega)$ [$\delta T_c/\delta \alpha^2$F($\omega$)] and $\alpha^2F(\omega)$ spectrum of H$_3$S ($Im\bar3m$), in the pressure range where the high-T$_c$ was measured (155-225 GPa). The calculations are done in the framework of the Migdal–Eliashberg theory. We find for this electron-phonon superconductor, a correlation between the maximums of $\delta T_c/\delta \alpha^2$F($\omega$) and $\alpha^2F(\omega)$ with its higher T$_c$. We corroborate this behavior in other electron-phonon superconductors by analyzing data available in the literature, which suggests its validity in this type of superconductors. The correlation observed could be considered as a theoretical tool that in an electron-phonon superconductor, allows describing qualitatively the proximity to its highest T$_c$, and determining the optimal physical conditions (pressure, alloying or doping concentration) that lead to the superconductor reaching its highest T$_c$ possible.

\end{abstract}



\maketitle

\section{Introduction}

From the linearized version of the Migdal–Eliashberg gap equations, it is possible to obtain the functional derivative of the superconducting transition temperature T$_c$ with respect to the electron–phonon coupling function $\alpha^2F(\omega)$, $\delta T_c/\delta \alpha^2$F($\omega$), from which the effectiveness of the phonons at different frequencies in building up the superconducting state can be evaluated~\cite{ba,ma}. The first numerical calculations of the $\delta T_c/\delta \alpha^2$F($\omega$) were performed by Bergmann and Rainer~\cite{Ber,Ber2} for isotropic superconductors. Their results showed that this function has a universal shape: It increases from zero at $\omega = 0$ to a maximum at $\sim 7k_BT_c$, and then decreases slowly to 0 as $\omega \rightarrow \infty$~\cite{mi}. This maximum is related to the importance of relaxation effects in the electron-phonon interaction as it enters superconductivity~\cite{ba}. The functional derivative describes the change in $\Delta$T$_c$ to a small change in the function $\Delta\alpha^2F(\omega)$ due to pressure or alloying~\cite{mi,mi2}
\begin{equation}\label{DT}
\Delta T_c = \int_0^{+\infty}d\omega\frac{\delta T_c}{\delta \alpha^2F(\omega)}\Delta \alpha^2F(\omega)
\end{equation}
Some researches~\cite{sp1,sp2,sp3,sp4,sp5} have studied the derivation and characteristics of the functional derivative for isotropic and anisotropic superconductors, and other few works~\cite{ba,mi,mi2,app1} have explored its application an important number of conventional superconductors, where the frequency in which the coupling between phonons and electrons is best for superconductivity was identified, for each case. Nicol and Carbotte~\cite{ni} shown that high-T$_c$ H$_3$S superconductor (203 K~\cite{dr}) is one of the most highly optimized among conventional superconductors, it through evaluation of its functional derivative, under the idea that when moving all the spectral weight in the $\alpha^2F(\omega)$ to a delta function placed at the maximum of the functional derivative ($\omega \sim 7k_BT_c$), one will obtain the highest T$_c$ possible from this spectrum. It is important to point out that the application of Migdal-Eliashberg theory, in this case, it was possible since the superconductivity mechanism in H$_3$S is the electron–phonon interaction~\cite{epi}.
Overall, the theoretical studies on new high-T$_c$ superconductors are aimed at fully understanding the superconducting mechanism for achieving the prediction of T$_c$ in new similar compounds.
In this paper, we report a study of the functional derivative of T$_c$ with $\alpha^2F(\omega)$ and $\alpha^2F(\omega)$ spectrum for H$_3$S under pressure. For this purpose, we calculate the $\delta T_c/\delta \alpha^2$F($\omega$) from the linearized version of the Migdal–Eliashberg gap equations and work with $\alpha^2F(\omega)$ spectra provided by a previous paper~\cite{ca}. Our results are compared with data available in the literature for other conventional superconductors (Nb$_3$Ge and Li (fcc)).

\section{Method of Calculation}

The functional derivative of T$_c$ with $\alpha^2F(\omega)$, $\delta T_c/\delta \alpha^2F(\omega)$, is determined by using, in a first step, the solution of the Linearized Migdal–Eliashberg equations (LMEE) on the imaginary axis~\cite{Ber,Ber2,Daa}($n,m$ are integers):
\begin{equation}\label{LMEE}
\rho\bar\Delta_n=\pi T \sum_m\left(\lambda_{nm}-\mu^*-\delta_{nm}\frac{|\tilde\omega_n|}{\pi T}\right)\bar\Delta_m,
\end{equation}
with
\begin{equation}
    \tilde\omega_n=i\omega_n+\pi T \sum_m \lambda_{nm}sgn(\omega_{m}),
\end{equation}
where at temperature $T$, $\omega_n=\pi T (2n-1)$ is the $n-th$ Matsubara frequency; and
\begin{equation}\label{GapEquation}
\bar{\Delta}_n=\frac{\tilde{\Delta}_n}{\rho+|\tilde{\omega}_n|},
\end{equation}
in terms of $\tilde\Delta_n$, the superconductor gaps. $\rho$ is a pair break parameter without physical meaning. The solution $T=T_c$ is getting at the limit when $\rho\rightarrow 0$.
The kernel $K_{nm}$ is given by, 
\begin{equation}
K_{nm}=\pi T \left[\lambda_{nm}-\mu^*-\delta_{nm}\frac{|\tilde\omega_n|}{\pi T}\right].
\end{equation}
Then a $\delta K_{nm}$ variation drives to change $\delta \rho$,
\begin{equation}
\delta \rho = \frac{\sum_{nm}\bar{\Delta}_n\delta K_{nm}\bar{\Delta}_m}{\sum_n\bar{\Delta}^2_n}
\end{equation}
where $\bar{\Delta}_n$ is the corresponding eigenvector to $\rho=0$ and $\delta K_{nm}$ is evaluated at $T=T_c$. Finally, functional derivative is calculated by
\begin{equation}
\frac{\delta T_c}{\partial\alpha^2F(\omega)}=-\frac{\frac{\delta\rho}{\delta\alpha^2F(\omega)}}{\left(\frac{\partial\rho}{\partial T}\right)_{T_c}},
\end{equation}
where to $\left(\frac{\partial\rho}{\partial T}\right)_{T_c}$ and $\frac{\delta\rho}{\delta\alpha^2F(\omega)}$ were used 
\begin{equation}
\left(\frac{\delta K_{nm}}{\delta T}\right)_{T_c}=\pi T_c\left[\left(\frac{\partial \lambda_{nm}}{\partial T}\right)_{T_c}-\delta_{nm}\sum_{m'}\left(\frac{\partial \lambda_{nm'}}{\partial T}\right)_{T_c}sgn(\omega_{m'}\omega_n)\right].
\end{equation}
and
\begin{equation}
\frac{\delta K_{nm}}{\delta \alpha^2F(\omega)}=\pi T_c\left[\frac{\delta \lambda_{nm}}{\delta \alpha^2F(\omega)}-\delta_{nm}\sum_{m'}\frac{\delta \lambda_{nm'}}{\delta \alpha^2F(\omega)}sgn(\omega_{m'}\omega_n)\right].
\end{equation}
The Eliashberg spectral function $\alpha^2$F($\omega$) calculated at different pressures were taken from a previous work~\cite{ca}, which were obtained in the pressure range where the high T$_c$ was measured (155-225 GPa)~\cite{dr}. The dynamic stability of H$_3$S $Im\bar3m$ in this pressure range was previously confirmed~\cite{ca}.

\section{Results and discussion} 

\begin{figure}[h!]
\centering
\includegraphics[width=0.8\textwidth]{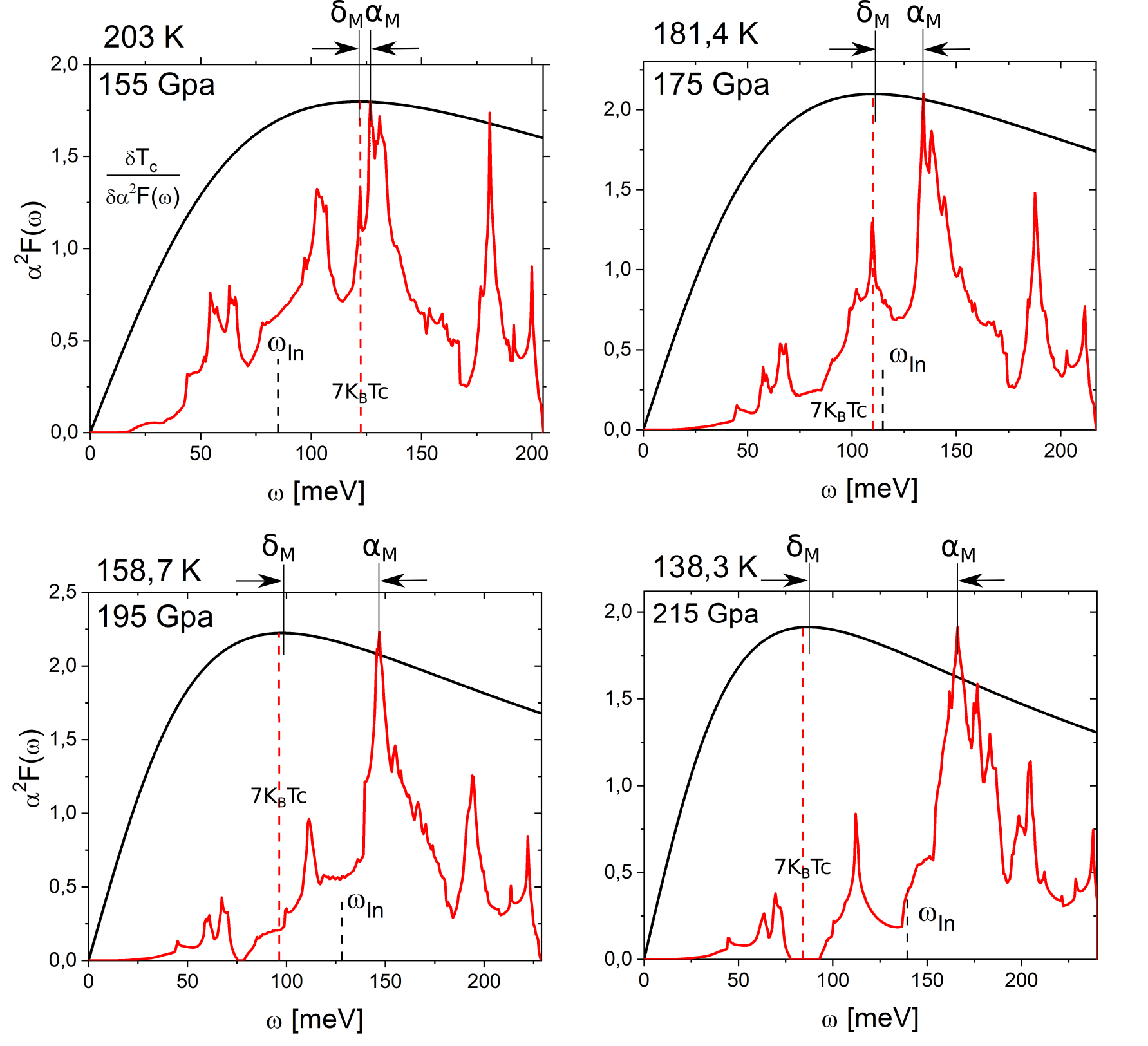} 
\caption{Pressure effects on the Eliashberg function $\alpha^2F(\omega)$ and functional derivative $\delta T_c/\delta \alpha^2$F($\omega$) for H$_3$S. The functional derivatives were scaled up to permitting comparison with the respective $\alpha^2$F($\omega$) spectrum. Each figure is labelled with its corresponding pressure and T$_c$. The arrows indicate the evolution of maximum peaks for $\alpha^2$F($\omega$) and $\delta T_c/\delta \alpha^2$F($\omega$), $\alpha_M$ and $\delta_{M}$ respectively. The frequencies $\omega_{ln}$ and $\omega=7k_{B}T_c$ are also calculated and indicated (dashed lines).}
\label{F1} 
\end{figure}

The Eliashberg spectral functions $\alpha^2$F($\omega$) and functional derivative $\delta T_c/\delta \alpha^2$F($\omega$) for H$_3$S calculated at different pressures are shown in Fig.~\ref{F1}. By and large, our $\alpha^2$F($\omega$) spectra are in good agreement with previous theoretical reports (at 155 and 200 GPa)~\cite{aa1,a1,a2,a3,a4}. 

It is observed in Fig.~\ref{F1} that pressure induces an almost rigid displacement of the $\alpha^2$F($\omega$) spectrum and its maximum peak ($\alpha_{M}$) towards higher energies. A considerable decrease in the area under the curve of this spectrum, mainly in the low-frequency region (10-90 meV), is also observed. For more details see reference~\cite{ca}.

For functional derivative, the increase in pressure induces a slight displacement of its maximum peak ($\delta_{M}$) towards lower energies. It has a sharper maximum for the sample with the lowest T$_c$ (138,26 K to 215 GPa). This maximum gets broader as T$_c$ goes up. In general, it is observed that the maximum is sharp for weak coupling and it gets broader as the coupling strengthens~\cite{ba}. It is also possible to observe that $\delta_{M}$ is located at $\omega \sim 7k_BT_c $ ($k_B$ is the Boltzman constant), although a slight difference grows with increasing pressure.

The T$_c$, $\delta_ {M} $ and $\alpha_ {M}$ values obtained at different pressures are shown in Table 1. T$_c$ values were calculated in a previous work~\cite{ca}, which are in good agreement with the experimental values~\cite{dr}.

\begin{table}[htbp]
\begin{center}
\begin{tabular*}{0.415\textwidth}{|c|c|c|c|c|}\hline\hline
P      &	T$_c$ 	&$\delta_{M}$	& $\alpha_M$ &	$\delta_{M}-\alpha_M$ \\
(GPa)  &  (K)     & (meV)       & (meV)      &  (meV)      \\\hline
155    &	203,000 &	121,7       &	126,8      &	5,1  \\
160    &	197,400 &	113,9       &	128,7      &	14,8 \\
165    &	191,388 &	115,5       &	132,1      &	16,6 \\
170    &	186,263 &	112,8       &	132,0      &	19,2 \\
175    &  181,438 &	110,2       &	134,3      &	24,1 \\
185    &	169,300 &	103,6       &	139,5      &	35,9 \\
195    &	158,670 &	98,00       &	147,1      &	49,1 \\
205    &	147,305 &	91,80       &	152,7      &	60,9 \\
215    &	138,258 &	86,90       &	166,1      &	79,2 \\
\hline\hline
\end{tabular*}
\caption{The T$_c$, $\delta_ {M} $ and $\alpha_ {M}$ values obtained at different pressures. $\alpha_M$ and $\delta_{M}$ are the maximum peaks of $\alpha^2$F($\omega$) and $\delta T_c/\delta \alpha^2$F($\omega$) respectively. T$_c$-values taken from reference~\cite{ca}.}
\label{T1}
\end{center} 
\end{table}

An important feature in Fig.~\ref{F1} and Table~\ref{T1} is observed. When comparing the values of $\delta_{M}$ and $\alpha_{M}$ energies, their proximity reveals to be correlated with the value of the T$_c$($\omega$). A small difference in $\delta_{M}-\alpha_M$ implies the system being closer to the optimal frequency ($\omega_{opt}= 7k_BT_c$), this is to a higher critical temperature, and vice versa. 

\begin{figure}[h!]
\centering
\includegraphics[width=1.0\textwidth]{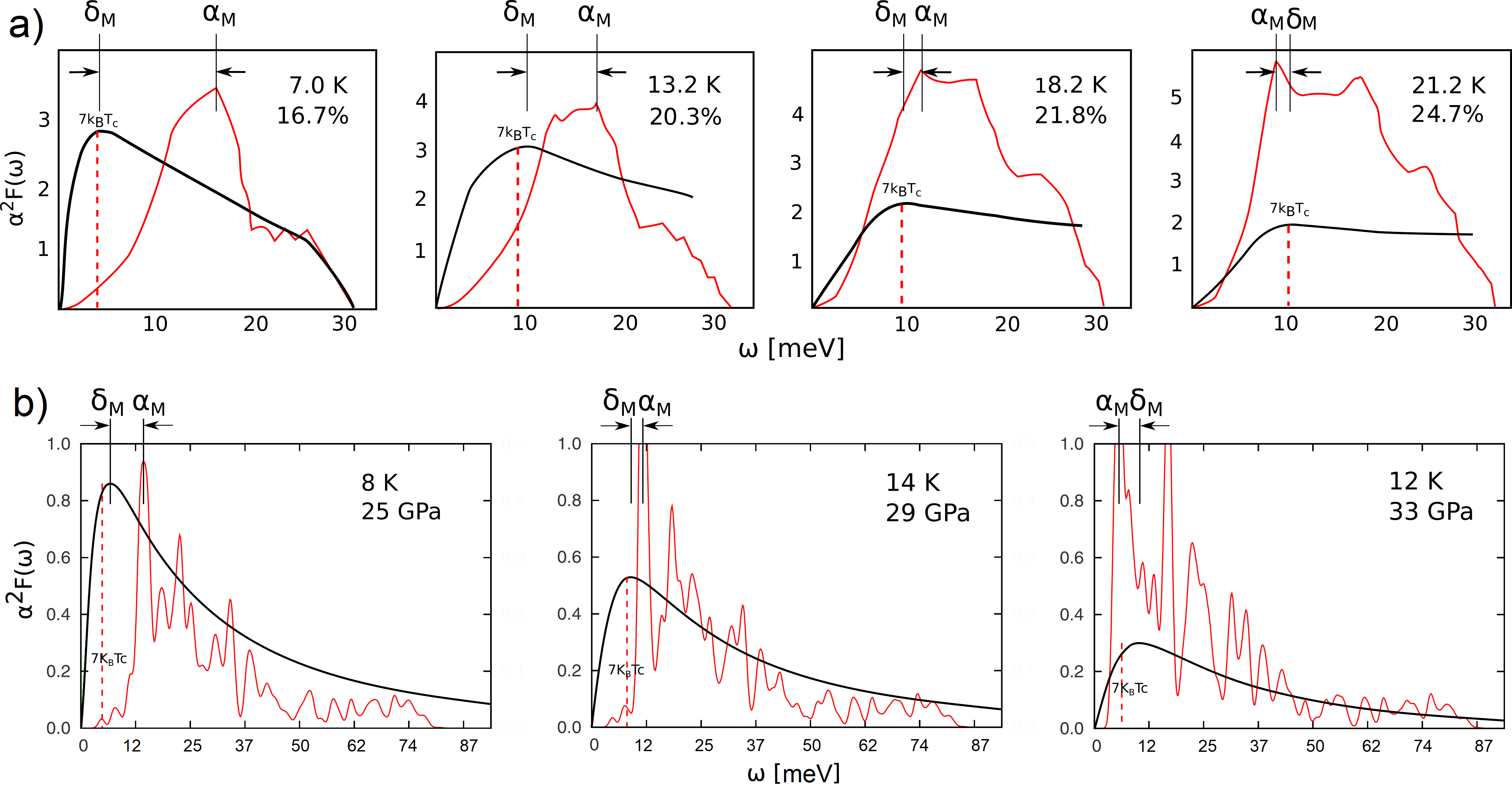} 
\caption{Eliashberg spectral function $\alpha^2F(\omega)$ and functional derivative $\delta T_c/\delta \alpha^2$F($\omega$) for Nb$_3$Ge and Li (fcc) given in references~\cite{ba,ma}. a) NbGe system to different Ge concentration (in \%). b) Pressure effects on Li (fcc). Each picture is labelled with its corresponding experimental T$_c$~\cite{Ki,Ki2,de}. The arrows indicate the evolution of the difference $\delta_{M}-\alpha_M$. The frequency $\omega_{opt}=7k_{B}T_c$ are also indicated (dashed lines).}
\label{F2} 
\end{figure}

Similar behavior can be observed in Nb$_3$Ge and Li (fcc) as reported by Baquero et al~\cite{ba} and Yao et al~\cite{ma}, respectively. Fig.~\ref{F2}a describes the evolution of the Nb-Ge system towards stochiometry and Fig.~\ref{F2}b shows the effects of pressure superconductor properties on Li. In both cases, it is observed that the closest proximity of $\delta_{M}$ and $\alpha_{M}$ occurs for the configuration in which the system has the highest T$_c$. However, for Li, the highest T$_c$ does not imply the largest broad of the maximum of $\delta T_c/\delta \alpha^2$F($\omega$), as it occurs in H$_3$S and Nb$_3$Ge. It is also possible to observe that in both systems $\delta_{M}$ and $\alpha_M$ are close at $\omega=7k_BT_c$ when the system has the highest T$_c$. According to our best knowledge, there are no other data available to extend the verification of this correlation.

These results are consistent with the idea of moving all the spectral weight in the $\alpha^2F(\omega)$ to a delta function placed at $\omega \sim 7k_BT_c$, which would then allow obtaining the highest T$_c$ of that system~\cite{ba,ni}. In this regard, Nicol and Carbotte~\cite{ni} proposed that the Allen-Dynes characteristic phonon frequency $\omega_{ln}$ ($\omega_{ln}=exp\{2/\lambda\}\int_0^{\infty}[ln(\omega)\alpha^2F(\omega)/\omega]d\omega$)\cite{ad} could be used as a close measure to this delta function, for H$_3$S. However, our results show $\omega_{ln}$ varies significantly with pressure ($\omega_{ln}$ increases with increasing pressure, see Fig.~\ref{F1}), which implies that possibly $\omega_{ln}$ is not a suitable reference parameter for such comparison. For H$_3$S, the variation of $\omega_{ln}$ versus pressure have been also reported by Akashi et al~\cite{Aka}.

Based on our results, we conclude from the behavior seen in these systems that electron-phonon superconductors will reach their highest possible T$_c$ ($\omega_{opt}=7k_{B}T_c$) when the maximum peaks of the $\alpha^2F(\omega)$ spectrum and the functional derivative of T$_c$ with respect to $\alpha^2F(\omega)$ manage to get as close as possible. 

This behavior could be considered as a theoretical tool that, in an electron-phonon superconductor, allows testing the degree of proximity to its highest T$_c$ possible for physical conditions defined. Furthermore, it could establish (through variations of spectral function $\alpha^2$F($\omega$)) the optimal physical conditions (pressure, alloying or doping concentration) that lead to the superconductor have the highest spectral density of the electron-phonon interaction $\alpha_M$ at the optimal vibrational frequency $\omega_{opt}$, that is to say, its highest T$_c$. 

\section{Conclusions}

In this work, we report a study of the functional derivative of T$_c$ with Eliashberg spectral function $\alpha^2F(\omega)$ [$\delta T_c/\delta \alpha^2$F($\omega$)] and the $\alpha^2F(\omega)$ spectrum for H$_3$S ($Im\bar3m$) under pressure (155-215 GPa). The $\delta T_c/\delta \alpha^2$F($\omega$) were calculated from the linearized version of the Migdal–Eliashberg gap equations. Our results were analyzed and compared with data available in the literature.

We found a correspondence between the maximum peaks of $\alpha^2$F($\omega$) spectrum ($\alpha_M$), $\delta T_c/\delta \alpha^2$F($\omega$) ($\delta_{M}$) and the highest T$_c$, which satisfies the idea of moving all the spectral weight in the $\alpha^2F(\omega)$ to a delta function placed at $\omega_{opt} = 7k_BT_c$ (optimal frequency vibrational) where the system will be optimized. We observed that the convergence of $\delta_{M}$ and $\alpha_{M}$ (as $\alpha^2$F($\omega$) varies) implies that the system is close to the higher T$_c$, and vice versa. This behavior was corroborated in other conventional superconductors by analyzing data available in the literature, suggesting its validity in electron-phonon superconductors. 

Based on our results, we suggest that this correlation could be a theoretical tool that, in an electron-phonon superconductor, allows testing the degree of proximity to its highest T$_c$ possible for physical conditions defined, furthermore could establish (through variations of spectral function $\alpha^2$F($\omega$)) the optimal physical conditions (pressure, alloying or doping concentration) that lead to the superconductor reaching its highest T$_c$ possible.

\section{Acknowledgments}

The author acknowledge to the CGSTIC at Cinvestav for providing HPC resource on the Hybrid Cluster Supercomputer {\em Xiuhcoatl} and to {\em Universidad del Rosario} for providing HPC resource on Cluster of {\em Laboratorio de Computación Avanzada}. J.C. and G.G. wish to recognize and thank Professor R. Baquero for his unconditional friendship and academic support.

\end{document}